\documentclass[12pt]{article}

\pdfoutput=1

\usepackage{graphics}
\usepackage{amssymb}
\usepackage{amsmath}
\usepackage{hyperref}
\usepackage{bbold}
\usepackage{tikz}

\newcommand{\beq}{\begin{equation}}

\newcommand{\eeq}{\end{equation}}
\newcommand{\bea}{\begin{eqnarray}}
\newcommand{\eea}{\end{eqnarray}}

\newenvironment{dedication}
    {\vspace{6ex}\begin{quotation}\begin{center}\begin{em}}
    {\par\end{em}\end{center}\end{quotation}}

\begin{document}

\begin{center}
${}$\\
%\vspace{100pt}
\vspace{50pt}
{ \Large \bf Nonperturbative Quantum Gravity
\\ \vspace{10pt} Unlocked Through Computation 
}

\vspace{36pt}

{\sl R. Loll}

\vspace{18pt}
{\footnotesize

Institute for Mathematics, Astrophysics and Particle Physics, Radboud University \\ 
Heyendaalseweg 135, 6525 AJ Nijmegen, The Netherlands.
%{email: r.loll@science.ru.nl}\\
}

\vspace{-36pt}
\end{center}
\begin{dedication}
This contribution is dedicated to Chris Isham, my former teacher and our all quantum gravity guru of old.
\end{dedication}
\vspace{12pt}
\begin{center}
{\bf Abstract}
\end{center}

\noindent 
Being able to perform explicit computations in a nonperturbative, Planckian regime is key to understanding quantum gravity as a fundamental
theory of gravity and spacetime. Rather than a variety of different \textit{approaches} to quantum gravity, 
what we primarily need is a gravitational analogue of the highly successful
lattice treatment of nonperturbative quantum chromodynamics. 

Unsurprisingly, however, lattice quantum gravity is not simple.
The crucial insight that has finally led to success is to \textit{build the dynamical and Lorentzian nature of 
spacetime into the lattices from the outset}. Lattice quantum gravity based on causal dynamical triangulations (CDT) puts this idea into practice
and is producing new and exciting physical results from numerical experiments.

This largely nontechnical account describes the challenges and achievements of modern lattice quantum gravity, which has opened an unprecedented
computational window on quantum spacetime in a Planckian regime and is reshaping our understanding of what it means to ``solve" quantum gravity. 
This methodology is well placed to unlock the physics of the early universe from first principles.  
Related topics discussed are the difference between lattice and discrete quantum gravity, and the role of spacetime emergence in the light of computational results.

\vspace{12pt}
\noindent

\newpage

\noindent{\large\bf Quantum gravity and the lack of computability}
\vspace{0.4cm}

\noindent \textit{Understanding quantum gravity}, the elusive fundamental quantum (field) theory underlying the classical theory of general relativity, obviously includes the 
ability to perform computations that quantify its physical properties, for example, in terms of the spectra of suitable quantum observables.  
Since gravity is pertur\-ba\-tively nonrenormalizable, neither perturbative nor effective quantum field theoretic methods are sufficient to reach such an understanding, 
and an approach beyond perturbation theory is called for. 

Historically, the failure of standard, perturbative\footnote{A perturbative quantization is based on splitting the metric field tensor $g$ into the constant Minkowski metric $\eta$
and a small perturbation $h$ according to $g_{\mu\nu}(x)\! =\! \eta_{\mu\nu}\! +\! h_{\mu\nu}(x)$.} tools, encapsulated in the famous two-loop computations of 
Goroff, Sagnotti \cite{Goroff1985} and van de Ven \cite{{vandeVen1991}}, and the failure of standard, nonperturbative tools, 
in the attempt to emulate the successes of lattice QCD\footnote{see \cite{Loll1998} for a review of early lattice quantum gravity}, 
led to a search for \textit{non}standard solutions of quantum gravity. The 1980s saw the beginnings of what many popular science publications continue to
call ``the two leading approaches to quantum gravity", superstring theory and loop quantum gravity. One of the few features these two formulations have in
common is a reliance on fundamental, one-dimensional excitations, which places them outside the conventional framework of quantum field theory. 
Whether the assumption of their strings or loops is
correct or indeed testable remains unclear, but, more to the point, in either approach one is still far from being able to perform any meaningful computations 
on distance scales near the Planck length ($1.6\times 10^{-35}\, m$), let alone make new physical predictions based on such computations.

Because of this lack of nonperturbative computational tools, many discussions and disputes about quantum gravity in recent decades have 
focused on formalisms rather than results and on general principles and concepts rather than concrete calculations of local dynamics. 
Long wish lists of problems that quantum gravity should solve -- if only we knew what it was -- have been compiled \cite{Armas2021}, largely without the benefit  
of numerical or other reality checks. Moreover, the weakness of the gravitational interactions makes it unlikely that we will accidentally encounter a new physical
phenomenon attributable to quantum gravity that could provide guidance on building a fundamental theory.

Long before addressing the challenge of experimental verification, which is due to the extreme scales involved, we therefore seem to face a very unsatisfactory 
situation: without a sufficiently stringent computational framework it is difficult to formu\-late objective criteria
for the validity and correctness of candidate theories, and even a requirement like internal consistency becomes a hazy notion. 
Invoking quali\-tative, ``intuitive" criteria like simplicity or beauty instead may be outright misleading, 
given that the searched-for theory describes an unknown physical regime far beyond classicality and the validity of perturbation theory, where 
many standard concepts of classical spacetime are not expected to apply. 

Another elephant in the room that deserves attention is a high degree of nonuniqueness, which is associated with a large number of free parameters and
other free choices that come with a particular candidate theory, and imply a lack of physical predictivity. As we have learned from string theory, the prime example
of a grand unified theory of all the interactions, including gravity, this correlates with the theory's
richness of ingredients, in this case, many unobserved fundamental excitations, supersymmetry and extra dimensions.
It raises the interesting question of how little in terms of ingredients we can get away with when constructing a theory of quantum gravity, to avoid such a scenario. 

\vspace{0.7cm}
\noindent{\large\bf Quantum gravity is not simple}
\vspace{0.4cm}

\noindent However, it is worth emphasizing that there is absolutely no reason to expect quantum gravity to be a \textit{simple} theory, even in the absence of exotic ingredients.
Let us begin by recalling the complicated structure of the \textit{classical} field theory of general relativity, whose basic field is a Lorentzian metric $g_{\mu\nu}(x)$. 
The local cur\-va\-ture properties of such a spacetime are encoded in its Riemann tensor, a quantity with 256 com\-po\-nents, and its
dynamics are described by a coupled set of nonlinear partial differential equations, whose exact solution is only known in very special cases \cite{Stephani2009}.
Highly refined and dedicated numerical methods are needed to extract the physical content of Einstein's equations whenever gra\-vity is strong, like in
the collision of black holes \cite{Baumgarte2021}. As already stressed above, the quantum theory has an analogous need for numerical tools. 

Another useful reference for estimating complexity and computability are the quantum field theories of the standard model of particle physics, 
and quantum chromodynamics (QCD) in particular, which, like gravity, has nonlinear classical field equations and a complicated gauge group action. 
Most relevant to this comparison is its nonperturbative sector, which has been investigated extensively with the help of powerful lattice methods, yielding quantitative 
results about the QCD spectrum not obtainable by other means \cite{Workman2022}. Likewise, quantum gravity is an inter\-acting, nonperturbative quantum field theory, 
but with an arguably even more complicated field content, dynamics and symmetry structure.
This strongly suggests that quantum gravity will \textit{not} be simple, computationally or otherwise, and certainly not simpler than QCD. 
In other words, hoping for a magical insight that dissolves the known structural features and difficulties of gravity to yield a quantum theory governed
by simple relations and dynamical outcomes seems highly unrealistic.

In further assessing what is and is not feasible in quantum gravity, we also need to examine the power and limitations of our most advanced theoretical and 
computational tools, and to what extent they have enabled us to quantitatively understand the nongravitational fundamental interactions and QCD in particular.
This provides useful benchmarks for what can realistically be achieved in terms of ``solving" quantum gravity.\footnote{``solving" is put in inverted commas
to indicate that it is not clear a priori what this notion entails in nonperturbative quantum gravity}
On top of this, we still need to take into account that these tools must be
adapted to the gravitational case, where spacetime is dynamical and not part of the fixed background structure.

Lastly, exact mathematical methods are unlikely
to solve quantum gra\-vity, because they are not even able to describe the renormalizable quantum field theo\-ries of particle physics. This message may be
underappreciated, since there have been many studies of quantum gravity-inspired toy models that are sufficiently simple to allow for an exact
treatment beyond perturbation theory (see e.g.\ \cite{Budd2022}). 
However, they are based on unphysical, simplifying assumptions, like reducing the spacetime dimension from four to two or three, or 
by postulating additional spacetime symmetries.\footnote{Potentially confusing for non-experts, such toy models sometimes run under the label ``quantum gravity"
without highlighting their limited character.} Since this removes exactly the features that make physical quantum gravity in four dimensions interesting and
difficult, it is not surprising that these models teach us very little about the full theory (see also \cite{Loll2022}, Sec.\ Q17 for further discussion and references).

\vspace{0.7cm}
\noindent{\large\bf Lattice quantum gravity reloaded}
\vspace{0.4cm}

\noindent Viewing quantum gravity through the lens of computability and the availability of suitable tools, and combining this with some of the lessons of the 
last 40 years of quantum gravity research strongly suggests a refocusing on \textit{nonperturbative computation as the key to progress}. 
Since technical and conceptual issues in quantum gravity tend to be 
closely intertwined, this will help to inform expectations of what the theory can deliver. 
Recognizing the nature and magni\-tude of the challenge of making quantum gravity computable 
should not be a deterrent, but allow us to take a realistic perspective on the effort and time frame needed. 

Fortunately this effort does not have to start from scratch, since a quantum field theoretic formulation of quantum gravity with a functioning, well-tested and 
nonperturbative computational framework is already available. This ``lattice quan\-tum gravity 2.0" is formulated in terms of Causal Dynamical Triangulations (CDT)
and has been developed over the last 25 years \cite{Ambjorn2012,Loll2019,Ambjorn2024}, building on pre\-vious developments (see \cite{Loll1998} for a review).
Unlike its lattice predecessors ``1.0" it has the dyna\-mi\-cal and Lorentzian nature of spacetime built 
into its construction from the outset. 
In a nutshell, this formulation has opened a computational window near the Planck scale where certain ``numerical experiments" a.k.a.\ Monte Carlo 
simulations of quantum gravity can be performed with the help of dynamical lattice methods. Geometric quantum observables can be measured, 
giving us for the first time a quantitative insight into the nature and properties of 
quantum spacetime and its dynamics in this nonperturbative regime. Numerous nontrivial lessons have been learned in the course of these developments, 
and some of the results already obtained have been totally unexpected from the point of view of the classical and perturbative theories. 

The research program of CDT quantum gravity is very much ongoing, and its achievements and future perspectives will be sketched below. 
How does it fit into the larger quantum gravity landscape? 
Alongside the dominant superstring and subdominant loop paradigms (when measured in terms of publications, grant moneys and media attention),
research on alternative approaches beyond perturbation theory has always continued.\footnote{A timeline of the main developments of nonperturbative
quantum gravity since 1980 can be found in \cite{Loll2022}, Secs.\ Q15 and Q22.}  In broad brushstrokes, one can distinguish between more or less conventional 
quantum field theoretic formulations, where also lattice quantum gravity belongs, and other approaches with a looser or unclear relation to
the concepts of quantum field theory.
In the latter category are also formulations that posit some form of fundamental discreteness of spacetime, usually
at the Planck scale, like the causal set approach \cite{Surya2019}. 

When it comes to comparing these different candidate theories, one is faced with the
usual conundrum that there are no quantities that can be meaningfully compared, because of the theories' incompleteness and a lack of effective computational tools in 
many of them. Comparing the various formalisms instead is not a particularly fruitful exercise, because of their vastly different starting points and choices of
ingredients, which do not have a direct physical inter\-pre\-tation in them\-selves. Examples of a successful comparison are
the spectral properties of selected observables measured in CDT\footnote{the spectral dimension of (quantum) spacetime and and its so-called 
volume profile, cf.\ \cite{Ambjorn2024}} and aspects of renormalization group flows \cite{Ambjorn2024b},
which can be reproduced by using functional renormalization group methods 
in so-called Asymptotic Safety \cite{Reuter2019}, an approach that combines perturbative and nonperturbative elements of quantum field theory. 
If and when other nonperturbative computational schemes become operative in the future, 
we will be able to formulate further quantitative criteria to assess the equivalence or
otherwise of the corresponding candidate theories of quantum gravity. 

\vspace{0.7cm}
\noindent{\large\bf Lattice quantum gravity is \textit{not} discrete quantum gravity}
\vspace{0.4cm}

\noindent In the search for a theory of nonperturbative quantum gravity, so-called funda\-men\-tally discrete approaches have had an enduring popularity among 
practitioners. Loosely speaking, the underlying idea is that spacetime should come in discrete units or ``bits", just like all matter is composed of elementary
quantum particles. These building blocks are usually assumed to be Planck length-sized, with or without individual shapes or other properties that are
usually guessed, either in analogy with systems on much larger scales or because of some other expediency. 
After adding a prescription for how the microscopic bits can interact or relate to each other, one then envisages that a large number of them will 
coalesce or self-organize dynamically in such a way that quantum gravity and spacetime ``emerge". 

If it could be realized, such a picture seems attractively simple: all we have to do is consider finite arrangements of some ``Lego blocks" of finite size, 
relying on our everyday intuition, which is much more attuned to working with natural 
num\-bers than with real ones. As an added bonus, if all physical quantities come in terms of some minimal, fundamental length unit, the infinities characteristic of 
quantum field theory and the ensuing need to renormalize them will simply disappear. Everyone can do quantum gravity! -- Not surprisingly, such a scenario is 
far too simple to be true or have anything much to do with gravity, for reasons that will be explained below.

For simplicity, let us ignore that any notion of fundamental discreteness requires an operational definition. 
Intuitive, classical ideas will be meaningless in a Planckian quantum regime that lacks a pre-existing spacetime, which in this scenario is expected 
to be generated dynamically and not put in by hand.  
Even assuming such a definition, we run into the problem that for any choice of building blocks and interaction rules at the Planck scale there will be an infinity 
of other choices that are equally well motivated. 
Imagine that a particular choice could be shown to lead to a viable candidate theory of quantum gravity, in the sense of reproducing one or more known features of general
relativity. (This is a necessary but not sufficient condition, and has not actually happened yet.) 
Then there will be many other choices that are equally viable in this sense, but which by construction are
\textit{different} theories at the same, Planckian scale, which after all is the primary habitat of quantum gravity. We conclude that formulations of quantum gravity 
based on the assumption of fundamental discreteness at the Planck scale have a structural problem because of their high degree of nonuniqueness,
with a corresponding lack of predictivity. 

Although the nonperturbative lattice formulation of quantum gravity also has some discrete features, they are not fundamental in nature, but part of a
regularization, where an unphysical lower (``ultraviolet") cutoff on the length of the lattice edges is employed at an intermediate stage of the calculation to ``tame" infinities. Subsequently, one takes a scaling limit
by sending this cutoff to zero while renormalizing coupling constants appropriately, a process which under favourable circumstances leads to an essentially 
unique\footnote{depending on at most a small number of parameters that have to be fixed by comparing with real-world experiment or observation} continuum 
quantum theory without infinities \cite{Montvay1994}. 
Importantly, by a mechanism called universality \cite{Goldenfeld1992}, the final theory does not depend on the details of how the regularization was set up,
like the shape of the building blocks or the detailed manner of their interaction. This provides the uniqueness mechanism that is missing in the
fundamental discreteness scenario.

Since lattice investigations are run on computers with finite processing power and storage capacity, the continuum limit of vanishing cutoff and infinitely fine
lattices cannot be reached in practice, but is extrapolated systematically from sequences of ever finer lattices. From a practical point of view, 
it implies that to extract universal results at a given
scale, say, the Planck length, one must use a lattice resolution that is significantly smaller than this scale to avoid that the measurements are
dominated by discretization artefacts. 

\vspace{0.7cm}
\noindent{\large\bf CDT and the challenges of lattice quantum gravity}
\vspace{0.4cm}

\noindent Let us introduce the basic principles and structural features that enter into the construction of modern lattice quantum gravity, based on
causal dynamical triangulations. Rather than an \textit{approach} to quantum gravity, distinguished by a specific choice of nonstandard ingredients (loops, 
strings, spin foams, causal sets or others \cite{Handbook2024}), it is a minimal nonperturbative quantum extension of
general relativity, using only standard principles from (lattice) quantum field theory \textit{adapted to accommodate the dynamical nature of spacetime}.

This adaptation, beyond the framework of relativistic quantum field theory on a fixed Minkowski space, did not just involve a few minor tweaks, but 
required solutions to long-standing problems that have hampered many approaches to quantum gravity: 
how to regularize and renormalize in a way that is compatible with diffeomorphism invariance, how to analytically continue (``Wick rotate") the path integral
to make it amenable to computation, how to deal with the conformal divergence of the resulting path integral \cite{Dasgupta2001}, and how to achieve 
unitarity \cite{Ambjorn2012,Loll2019}.

The dynamical principle at the heart of the quantum theory is the usual Feynman path integral, which in the gravitational case implements 
the quantum superposition of curved spacetime geometries $g$, schematically written as the functional integral
\begin{equation}
Z=\int_{\cal G} {\cal D}g\; \mathrm{e}^{i S[g]},
\label{pathint}
\end{equation}
where each geometry $g\!\in\! {\cal G}$ is weighed by a complex phase factor depending on the gravitational action $S[g]$. The expression
(\ref{pathint}) is entirely formal and needs to be accompanied by explicit definitions of the nonlinear configuration space $\cal G$ and its parametrization, 
the measure ${\cal D}g$ and a prescription for how to compute $Z$ nonperturbatively, without resorting to a perturbative linearization of $\cal G$ around a
solution of the classical Einstein equations.   

However, even after making these specifications, $Z$ will be ill-defined and infinite, due to quantum field theoretic divergences that need to be
renormalized. This has nothing to do with gravity, but also happens for a scalar field theory, say. 
At this point, it is natural to invoke a lattice regularization and renormalization to evaluate the path integral nonperturbatively, following the
highly successful example of lattice QCD. Such a strategy was suggested early on in the history of lattice gravity, 
using various classical gauge-theoretic (re-)formulations of gravity
as a starting point (see e.g.\ \cite{Smolin1978}). It was pursued for a number of years, including numerical lattice implementations, but 
remained unsuccessful and
inconclusive \cite{Loll1998}, as already mentioned earlier.

Even in hindsight it is difficult to pinpoint which of the shortcomings of ``lattice gravity 1.0" 
contributed most to this negative outcome, but a prime culprit was the
use of fixed, hypercubic lattices, on which the gravitational holonomy variables were placed. 
The problem is that the diffeomorphisms\footnote{smooth invertible 1-to-1 maps of the
underlying manifold to itself}, which form the invariance group of general relativity, do not act on such lattices and the naively discretized
continuum fields defined on them. It implies that
the corresponding gauge group action cannot be ``factored out" in a controlled way and the lattice fields do not properly represent the physical gravitational 
degrees of freedom.

Another major problem is that the Monte Carlo (MC) techniques used to evaluate lattice-regularized path integrals require a \textit{Euclidean}
quantum field theory. For a theory on Minkowski space this can be obtained by an analytic continuation from 
real to imaginary time, which under suitable conditions converts the complex phases $\mathrm{exp}(iS)$ in the path integral (\ref{pathint}) to real 
Boltzmann factors $\mathrm{exp}(-S^{\mathrm eu})$,
as needed in the MC simulations\footnote{A set of conditions a Euclidean (lattice) quantum field theory must satisfy to allow for a rotation back 
to Lorentzian signature is discussed in \cite{Montvay1994}.}, where $S^{\mathrm eu}$ denotes the action of the Euclidean theory. 
The problem in quantum gravity beyond perturbation theory is that spacetime and 
therefore also time are dynamical. For arbitrary curved spacetimes
there is no distinguished choice of time and moreover the time dependence of the metric field tensor can be arbitrarily complicated. 
As a consequence, no Wick rotation for metrics $g_{\mu\nu}(x)$ is known that achieves the required conversion of the phase factors. 

This has motivated many researchers to consider a different and a priori unrelated theory, so-called Euclidean quantum gravity, 
which is defined by a real ``path integral" (a.k.a.\ a partition function) over Riemannian instead of Lorentzian geometries, possessing no notion of time or causality.
Prior to CDT, all attempts at lattice quantum gravity were of this type, including gauge-theoretic approaches, quantum Regge calculus \cite{Williams2009} and
Euclidean dynamical triangulations \cite{Ambjorn1997}. 
Even if one could make sense of these path integrals, which has proven very challenging, it is unclear what, if anything they have to do with the physical, 
Lorentzian theory of quantum gravity.

\vspace{0.7cm}
\noindent{\large\bf CDT: lattices going dynamical}
\vspace{0.4cm}

\noindent To cut a long story short, from its inception \cite{Ambjorn1998} to today, CDT quantum gra\-vity takes both the dynamical \textit{and} 
the causal, Lorentzian nature of spacetime into account by building them into the structure of the lattices from the outset. In other words, the functionalities of the
lattice as a tool for regularizing the infinities of the quantum field theory have been adapted to match the physical content and 
symmetry structure of gravity, which are significantly different from those of gauge field theories. The success of this ansatz until now shows that this is
a fruitful strategy.

The dynamical lattices of CDT, which represent distinct curved spacetimes in a regularized version of the continuum path
integral (\ref{pathint}), play the same role for gravity as the lattice representation of QCD field configurations in terms of holonomy variables due to
Wilson \cite{Wilson1974} does for the strong interactions. 
The beauty and power of the latter lies in the fact that the $SU(3)$-gauge transformations placed at the lattice vertices have a
well-defined action on the group-valued holonomy variables on the lattice edges, which yields an \textit{exact} notion of gauge-invariant lattice field configurations and
observables.    

The curved geometry of the CDT lattices is defined by the geodesic edge lengths of their simplicial, four-dimensional Minkowskian building blocks\footnote{since 
by construction all spacelike edges 
and all timelike edges  have the same length, there are only two types of geometrically distinct simplicial building blocks, up to time reflection} 
and the way in which
these four-simplices\footnote{four-simplices are the four-dimensional analogues of two-dimensional triangles and three-dimensional tetrahedra} 
are glued together pairwise to obtain a triangulation, i.e.\ a piecewise flat spacetime manifold \cite{Ambjorn2001}. 
The length and gluing data are geometric in nature, but in order to locate the corresponding four-simplices inside a tri\-an\-gu\-lation the simplices 
need to be numbered or ``labelled". This discrete labelling is arbitrary and unphysical in the sense that no observables can depend on it. 
The associated relabelling invariance may be thought of as an analogue of the coordinate- or diffeomorphism-invariance of general relativity,
but unlike the latter is easily taken into account when evaluating the path integral. 

Similar to what happens in Wilson's formulation of lattice QCD, CDT quantum gravity therefore has
an exact notion of gauge-invariance, despite the presence of a lattice cutoff. In this formulation, unlabelled triangulations represent manifestly coordinate-invariant 
spacetime geometries. This remarkable property is due to how the discrete gluings capture the local curvature degrees of freedom 
of the regularized spacetimes, without referring to any continuously varying metric variables. Compared to what is done in standard lattice field theory, 
making the lattice itself dynamical is key to intrinsically combining the powerful idea of approximating spacetime by a lattice with the dynamical 
character of spacetime in gravity. 

\vspace{0.7cm}
\noindent{\large\bf CDT: lattices going causal}
\vspace{0.4cm}

\noindent The idea of using dynamical, triangulated lattices has its origin in two dimensions \cite{David1984}, more precisely, 
the search for a nonperturbative description of the
dynamics of two-dimensional world sheets in bosonic string theory. In due course, its application to intrinsic, embedding-inde\-pen\-dent curved geometries in four
dimensions was considered, in an attempt to find a theory of Euclidean quantum gravity from a nonperturbative, regularized Euclidean path integral, of the kind
already mentioned above.
Claims of the presence of a second-order phase transition \cite{Agishtein1992} in the corresponding lattice quantum gravity model, 
so-called Euclidean dynamical triangulations (EDT)
or DT for short, signalling the possible
existence of a continuum limit, generated much attention at the time, but were later shown to be erroneous.\footnote{summaries of the set-up and results 
of EDT in four dimensions and further references can be found in \cite{Loll1998,Ambjorn2024,Ambjorn2024a}}.

By contrast, CDT quantum gravity has two decisive new elements, which lead to much more interesting outcomes, including the presence of second-order phase
transi\-tions \cite{Ambjorn2011,Coumbe2015} and the emergence of a macroscopic quantum spacetime with 
de Sitter properties \cite{Ambjorn2004,Ambjorn2007}, to be discussed further below. The first novel feature, compared to EDT, is the use of a path integral
over triangulated lattices which represent Lorentzian spacetimes rather than Riemannian spaces. Accordingly, one chooses Minkowskian
instead of Euclidean four-simplices as elementary lattice building blocks and associated gluing rules, which ensure that each triangulation contributing to the
path integral has a well-defined causal (or lightcone) structure globally \cite{Ambjorn2001}. 
Like in general relativity, each such spacetime consists of an ordered sequence of spatial slices representing 
moments in time. The letter ``C" in ``CDT" stands exactly for this causal ordering. 

However, formulating a lattice version of the physical, Lorentzian path integral (\ref{pathint}) is by itself not enough to achieve a
breakthrough, since this form cannot be used as a direct input for Monte Carlo simulations, as pointed out earlier. The crucial missing element is an 
analytic continuation of the complex path integral to a real partition function. Remarkably, this is also available in CDT and defines its second novel feature
as follows. It was already mentioned
that there are two different length assignments to the lattice edges, depending on whether they are space- or timelike. More precisely, 
spacelike edges have a squared\footnote{in Lorentzian signature it is more convenient to work with \textit{squared} lengths} 
length $\ell_s^2\! =\! a^2$ and timelike edges have $\ell_t^2\! =\! -\alpha a^2$, where $a$ is the so-called lattice spacing, i.e.\ 
the ultraviolet length cutoff that will be sent to zero eventually, and $\alpha\! >\! 0$ is a fixed positive constant, which from a classical point of view can be chosen arbitrarily. 
It has been shown that an analytic continuation of $\alpha$ to $-\alpha$ through the lower-half complex plane converts the path integral to a real
partition function of the correct functional form \cite{Ambjorn2001}, and makes it amenable to MC simulations. This is highly significant, since no analogous
prescription is known in a continuum formulation based on metric fields $g_{\mu\nu}$. It opens the door to a quantitative exploration of the
nonperturbative gravitational path integral.  

\vspace{0.7cm}
\noindent{\large\bf Emergence: aspirations and reality}
\vspace{0.4cm}

\noindent Independent of whether one follows the path of lattice quantum gravity, of a fundamentally discrete model, or of ``something totally different",
i.e.\ a conjectural theory with no resemblance to gravity as we know it at (sub-)Planckian scales\footnote{sub-Planckian means length scales even
smaller than the Planck length, sometimes also called trans-Planckian}, one needs to show that its predictions are compatible with those of
general rela\-ti\-vity in physical situations where quantum effects are negligible, usually at large length scales and/or low energies.
This turns out to be a very challenging task. 

By construction, the formulations and variables used to describe the classical and nonperturbative quantum regimes 
are very different, with smooth classical tensor fields like the metric $g_{\mu\nu}(x)$ playing no role in the latter.
More importantly, this will also be reflected in different observables and different ways in which diffeomorphism invariance is implemented, partly because
of the absence of an a priori background geometry at the Planck scale. 

The recovery of classical properties of gravity is a known difficulty of nonperturbative formulations. It is sometimes called the problem of the classical limit, but especially 
for the recovery of a classical spacetime, the term ``emergence" is invoked frequently. For clarification, since this notion is sometimes used in a very loose
sense, emergence here refers to (new) macroscopic properties arising from the collective behaviour of a large number of microscopic constituents. In the
case of quantum gravity, the question is whether and how (sub-)Planckian building blocks or other ingredients and their interactions can on
much larger scales give rise to an entity that
resembles a four-dimensional extended spacetime, as well as to gravitational dynamics as we know it from general relativity.\footnote{this excludes
analogue gravity models \cite{Barcelo2011}, which usually focus on recovering an effective Lorentzian spacetime, rather than a full-fledged theory of gravity}   

Emergence as an aspirational concept, without asking for a quantitative correlate, is seemingly straightforward: 
if we assume the presence of some incarnation of a Planckian ``quantum spacetime foam", 
a process of emergence arguably \textit{must} exist to take us from there to the classical theory. One might even dream of a
universal emergence mechanism, whereby a wide range of microscopic ingredients (``at the Planck scale, anything goes")
inevitably leads back to a nice classical spacetime like a Minkowski or de Sitter space.\footnote{as noted earlier, different Planckian ingredients will typically be
associated with different \textit{quantum} theories, but the present argument focuses on the emergence of \textit{classical} structures} 

A majority of formulations uses some variant of the path integral as their dyna\-mical principle, where
``emergence" comes about through a superposition of amplitudes.
Since there are currently no efficient com\-pu\-tational methods to evaluate the complex path integral (\ref{pathint}), one must rely on
a suitable analytic continuation like in CDT, or work with a real, Euclidean partition function or state sum. 
Fortunately, there is already a significant body of work where such systems have been analyzed quantitatively,
providing a much-needed reality check on the role of emergence. They 
include lattice quantum gravity, statistical models of random geometry, toy models in lower dimensions, 
and some discrete quantum gravity formulations like causal sets,
to the extent they allow a modicum of computational control. 
The overall conclusion is simple and largely independent of the details of the individual models: generically, \textit{nothing emerges}, or at least
nothing that has any obvious relation to general relativity or spacetime. 

This negative outcome can have various origins.
(i) Not enough is put in. Recall that classical gravity has a very complex local curvature structure and dynamics; if
the choice of microscopic ingredients and interactions is too minimalist and does not capture the potentiality of these rich classical structures,
they simply will not emerge -- one cannot get something for nothing. (ii) Too much is put in. If the set of configurations that is summed or integrated over 
in the path integral is too large, the resulting infinities cannot be renormalized with standard methods, and nothing emerges either.
The folklore that ``one should sum over everything in the path integral" simply does not make sense in nonperturbative quantum gravity. 

(iii) Even if (i) and (ii) are evaded, there are generic mechanisms which lead to a domination of the superposition by configurations that have nothing to do
with any recognizable space or spacetime macroscopically, no matter how they are weighed or coarse-grained. 
The point to appreciate here is that very large quantum fluctuations are present in the nonperturbative Planckian regime, which generically do not 
cancel each other out to lead to a quasi-classical space ``on average", i.e.\ in the sense of expectation values. Instead, even the dimensionality
of space can become dynamical, as is illustrated by the spectral dimension \cite{Ambjorn2005}. 

An infamous pathological mechanism of this type found in Euclidean
DT quantum gravity is \textit{polymerization}, whereby building blocks preferably arrange them\-selves\footnote{this is an ``entropic" effect, in as much as
there are many more ways for the building blocks to form distinct branched polymers than other, less degenerate macroscopic structures \cite{Ambjorn2010}} 
into a so-called branched polymer, with Hausdorff
dimension 2 and spectral dimension 4/3 in the limit of vanishing UV-cutoff, independent of the microscopic dimensionality $d$ of the building blocks(!), as
long as  $d > 2$.
A similar effect is also present in models other than DT (see \cite{Loll2022}, Sec.\ Q28 for further discussion and references). 

In spite of these difficulties, the idea of emergence is not doomed, since there is a known cure for both polymerization and crumpling \cite{Ambjorn1997},
another generic pathology. Judging by the results obtained, the crucial insight is to require path integral configurations to carry a well-defined causal structure, 
as one does in CDT.\footnote{note
that this causal structure is of course not fixed, but quantum-fluctuates alongside other aspects of the spacetime geometry} 
This has led to the first genuine instance of emergence in nonperturbative quantum gravity. More specifically,
there is strong, quantitative evidence from CDT lattice gravity for the dynamical generation of a quantum spacetime with properties that on sufficiently 
large scales match those of a (semi-)classical de Sitter space, in terms of dimensionality \cite{Ambjorn2004, Ambjorn2005a}, shape and its
quantum fluctuations \cite{Ambjorn2007,Ambjorn2008} and 
average curvature \cite{Klitgaard2020} (see also the reviews \cite{Ambjorn2012,Loll2019,Ambjorn2024}). 

Before taking a closer look at the nature of these results, and how it reflects both the nonperturbative physics and the corresponding toolbox, we can already
draw an important conclusion from the discussion above. It is not so much that achieving emergence is subtle and difficult, which is certainly true, but 
that nonperturbative computational tools have been absolutely essential in informing our current understanding of this phenomenon. 
It is impossible to guess the dynamical content of a given gravitational path integral or partition function without being able to evaluate it explicitly.
For example, the fact that microscopic four-dimensional building blocks do not generically give rise to macroscopic four-dimensional spaces 
in a continuum limit runs counter to any (semi-)classical or perturbative intuition and to many practitioners came as a great surprise. 
However, it merely illustrates the need for reality checks in the form of numerical experiments, to rein in
our often speculative ideas about quantum gravity and canalize them in the right direction.

\vspace{0.7cm}
\noindent{\large\bf Lattice quantum gravity: unlocking the early universe}
\vspace{0.4cm}

\noindent As argued above, quantum gravity is not simple and we should not expect it to be. It is significantly more complex than and structurally different 
from nonabelian gauge field theory, 
and one of the toughest problems theoretical physicists have set themselves. Keeping in mind the
time it took to observe gravitational waves -- a key prediction of the \textit{classical} theory -- and to get a grip on computing their waveforms, we should also not be 
surprised that the time scale for progress in quantum gravity is long. Fundamental quantum gravity shares the need to go beyond perturbation theory
with the nonperturbative sectors of general relativity and QCD, and the same essential need for effective computational tools to tackle this sector directly. 
In addition, taking into
account the lack of experimental guidance in quantum gravity, the case for computation as \textit{the} key to progress in understanding the theory is overwhelming. 

The go-to methodology in nonperturbative quantum field theory is a lattice regularization, but this did not succeed initially due to the static and Euclidean nature of
the lattices employed, as mentioned earlier. Since these early days, coming up with solutions to these issues and testing their viability has been a continuous, collective effort, 
culminating in a fully functional lattice formulation ``2.0" of quantum gravity. It is based on regularizing the gravitational path integral\footnote{the focus here is on
pure gravity, but the formalism allows for a straightforward coupling to matter fields if desired, see e.g.\  \cite{Loll2019} for further discussion} in terms of CDT, 
and has opened a measurement window on the terra incognita of Planck-scale physics. 

Like real experiments, also Monte Carlo experiments\footnote{Markov chain MC is the method of choice; quantum computing and machine learning
tools have been considered, but currently do not offer significant computational improvements \cite{privcomm}} are constrained by the resources available,
including computing power, storage capacity and efficiency of the algorithms (see e.g.\ \cite{Brunekreef2023}), and all measurement data are subject to 
statistical and systematic errors. 
Just like in lattice QCD, it requires ingenuity to construct observables and set up numerical experiments that yield reliable results for the available lattice sizes, 
which in typical simulations are of the order of $10^5$--$10^6$ simplicial building blocks. 

It should be emphasized that being able to compute \textit{anything} in nonperturbative quantum gra\-vity is an unprecedented situation.
It allows us to redirect focus away from formal matters and what the theory \textit{should} be like to the actual physical content of the theory and what it is able to deliver. 
Considering the status quo of lattice gauge theory and the decades of dedicated work that got us there, it is clear that 
in lattice quantum gravity much still lies ahead and that it will neither be easy nor happen overnight. 
Unlike QCD, quantum gravity is still at a much more exploratory stage, with a primary focus on finding new observables that can capture the physics
of the largely unknown nonperturbative regime. 
Further computational optimization will clearly be important, but will be tied to completely different
physical questions and observables than in nongravitational lattice theories. 

The lattice breakthrough allows us to also address a range of conceptual issues in quantum gravity within a concrete computational
framework, rather than based on abstract reasoning alone. Examples are the roles of time, causality, unitarity, topology change and spacetime symmetries at the Planck scale, 
some of which have already been clarified (see e.g.\ \cite{Loll2019}). Any question one poses has to be formulated as an operationally well-defined 
experiment that can be conducted in the accessible range of lattice parameters. ``Operationally well-defined" means among other things
that the observable whose eigenvalues are being measured must be labelling-invariant, which typically implies that it is of the form of a nonlocal
spacetime average (see \cite{Loll2022}, Sec.\ Q29, for further discussion of observables). Note that this requirement is not met for 
many semiclassically or perturbatively formulated questions that quantum gravity is expected to provide answers to, including the
black hole information loss problem \cite{Mathur2009}.

This situation is qualitatively different from the classical one, where reference systems in the form of local coordinate charts always exist, although they may be
nonunique and unphysical. In the nonperturbative realm no such coordinate or other reference systems exist. They also cannot be introduced by hand, in the
form of equipotential surfaces of some scalar fields \cite{Ambjorn2021} or by adding boundaries, say, because these will be subject to the same quantum 
fluctuations that prevent the existence of useful coordinate systems in the first place. In other words, the unfamiliar nonlocal character of observables is not due to some
shortcoming of the chosen formulation, but is an intrinsic feature of Planckian physics.  

The observables investigated in CDT lattice gravity so far provide concrete insights into the type of results one will be able to derive. They will be quantitative, but
obviously not of an analytical nature. One could wonder whether we will ever be able to develop an analytical description of this 
nonperturbative regime. Given the presumed strongly interacting character of gravity at the Planck scale, this seems exceedingly unlikely, but in
absence of no-go theorems in nonperturbative quantum gravity it cannot be excluded in principle. A more likely scenario is that we can
theoretically model selected aspects of the theory, based on the input from measurements. A good example are measurements of the correlator of
spatial three-volumes in the de Sitter phase of CDT lattice gravity, which have been used to reverse-engineer an effective cosmological action for the
scale factor \cite{Ambjorn2014}. Other quantities one can hope to extract in a (near-)Planckian regime are universal parameters
associated with the scaling behaviour of specific observables. The already mentioned spectral and Hausdorff dimensions \cite{Ambjorn2005,Ambjorn2005a} 
are of this type, and coefficients
characterizing the fall-off behaviour of diffeomorphism-invariant two-point functions \cite{Duin2024} would be another example.

The recent developments and results on lattices reshape our expectations of what quantum gravity is about and what it means to ``solve" it, 
i.e.\ what we may learn about it in the foreseeable future with the help of our best computational and theoretical tools. 
Given that it has already been shown that an extended quantum spacetime with \textit{some} de Sitter-like properties is generated dynamically in lattice gravity,
a tantalizing goal is to try to connect this to early-universe physics \cite{Glaser2017}, where the background spacetime for quantum fluctuations is usually 
\textit{assumed} to resemble a de Sitter universe. It would be spectacular if one could show that this assumption can be justified (or possibly
corrected) from first principles. This still requires highly nontrivial investigations, e.g.\ of the extent to which homogeneity and isotropy are present or 
``emerge" on larger scales \cite{Loll2023}, and an analysis of local quantum fluctuations and their correlators,
which are the subject of ongoing research. Importantly, there is a clear and concrete path forward, and computation is bound to unlock 
even more of quantum gravity's nonperturbative secrets.

\end{document}